\documentstyle[preprint,prl,aps,epsf]{revtex} 
\input epsf
\newcommand{\bb}{\begin{equation}}
\newcommand{\en}{\end{equation}}

\begin{document}
\draft
\title{Electrostatic Attraction of Coupled Wigner Crystals: Finite Temperature
Effects}
\author{A. W. C. Lau$^{1}$ and P. Pincus$^{1,2}$}
\address{$^{1}$ Department of Physics and $^{2}$ Department of Materials, 
University of California Santa Barbara, CA 93106--9530}
\author{Dov Levine}
\address{Physics Department, Technion-Israel Institute of Technology,
Haifa 32000, Israel}
\author{H.A. Fertig}
\address{Department of Physics and Astronomy, University of
Kentucky, Lexington, KY 40506-0055}
\date{\today}
\maketitle
\begin{abstract}
In this paper, we present a unified physical picture for the 
electrostatic attraction between two coupled planar Wigner crystals 
at finite (but below their melting) temperature. At very low temperatures, 
we find a new regime where the attraction, arising from the 
long-wavelength excitation of the plasmon mode, scales with 
the interplanar distance $d$ as $d^{-2}$.  
At higher temperatures, our calculation agrees with known results.  
Furthermore, we analyze the temperature dependence 
of the short-ranged attraction arising from ``structural'' correlations 
and argue that thermal fluctuations drastically reduce the strength of 
this attraction.
\end{abstract}
\narrowtext

\section{Introduction}

Electrostatic interactions play an important role 
in a system of charged macroions in an aqueous solution of
neutralizing counterions\cite{nature}. 
The macroions may be charged membranes, stiff polyelectrolytes such 
as DNA, or charged colloidal particles.  Recently, 
there has been a great interest in understanding the
attraction arising from correlations between highly-charged macroions as 
evidenced in experiments\cite{attractionE} and in
simulations\cite{attractionS}.  This attraction cannot be 
explained by the standard Poisson-Boltzmann (PB) treatment,
even for an idealized system of two highly charged planar 
surfaces, with counterions distributed between them, since PB, 
being a mean-field theory, neglects 
correlations.  Indeed, it has been proven recently that
PB theory predicts only repulsions between 
two likely-charged objects\cite{thm}.
Recall that the PB solution\cite{nature} for a single charged surface 
with charge density $en$ -- where $e$ is the elementary charge and 
$n$ the areal density -- immersed in a solution of neutralizing counterions
of valence $Z$, predicts a length scale $\lambda_{GC} 
= 1/(2\pi l_B Z n$) (where 
$l_B \equiv \frac{e^{2}}{\epsilon k_{B}T} \approx 7\,$\AA$\,\,$
is the Bjerrum length below which electrostatics dominates 
the thermal energy in an aqueous solution of dielectric constant 
$\epsilon = 80$ ($H_{2}O$), $k_B$ is the Boltzmann constant, 
and $T$ is the temperature.)  Physically, this Gouy-Chapman 
length $\lambda_{GC}$ defines a sheath near the charged surface
within which most of the counterions are confined\cite{manning}.  
For a moderately charged surface of $n \sim 1/100\,$\AA$^{-2}$, 
$\lambda_{GC}$ is of the order of few angstroms, and for highly charged 
surfaces and multivalent counterions $Z > 1$, we have $\lambda_{GC} < l_B$, 
signaling the breakdown of PB theory.  In this 
limit, fluctuations and correlations about the mean-field potential
become so large that the solution to the 
PB equation is no longer valid\cite{roland}.

To account for the attraction arising from correlations, two distinct
approaches have been proposed.  The first approach, based on charge 
fluctuations, treats the ``condensed'' counterion fluctuations 
in the Gaussian approximation.  This theory predicts a 
long-ranged attraction which vanishes as $T \rightarrow 0$\cite{charge}. 
In the other approach based on ``structural'' correlations 
first proposed by Rouzina and 
Bloomfield \cite{rb}, the attraction comes from the 
ground state configuration of the ``condensed'' counterions.  Indeed, 
at low temperature, the ``condensed'' counterions crystallize on 
the charged surface to form a 2D Wigner crystal.
When brought together, the counterions of two Wigner crystals 
correlate themselves to minimize the electrostatic energy.
These staggered Wigner crystals attract each other via
a short-ranged force that is strongest at $T=0$.  Although 
the physical origin of the attraction is clear in each approach,
the relationship between them remains somewhat obscure and their results 
in the $T \rightarrow 0$ limit are somewhat contradictory.
Therefore, it is desirable to formulate a unified approach 
which captures the physics of both mechanisms.

\begin{figure}
\vfil
\epsfxsize=6.0in
\epsfysize=2.25in
\centerline{\epsfbox{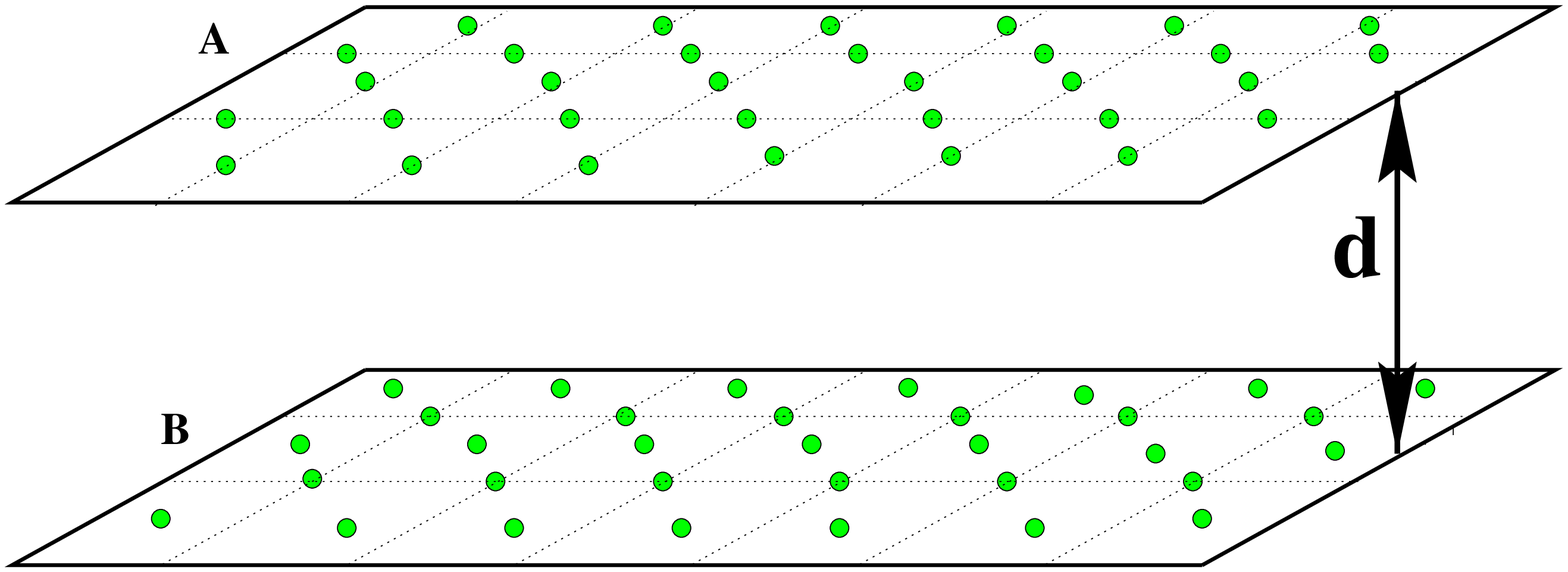}}
\vfil
\caption{A schematic picture of two staggered Wigner 
crystals formed by the ``condensed''
counterions at very low temperatures.}
\label{wigner}
\end{figure}

To this end, we attempt in this paper to develop a detailed 
physical picture of the electrostatic interaction at finite 
temperature between two planar Wigner crystals
in the strong Coulomb coupling limit.  Since correlation effects are 
essentially two-dimensional, we consider a model system composed 
of two uniformly charged planes a distance $d$ apart, 
each having a charge density $en$.  Confined on the surfaces 
are negative point-like mobile charges of magnitude $e$.  
In order to understand 
correlation effects that are not captured by PB theory, we assume
that the charges form a system of interacting Wigner crystals 
(see Fig. \ref{wigner}).  In particular, we compute the electrostatic 
attraction between the two layers by explicitly taking into account both 
{\em correlated fluctuations} and {\em ``structural'' correlations}.  
(By {\em ``structural'' correlations}, we mean the residual ground state
spatial correlations which remain at finite temperature.)
In the former case, we obtain a long-ranged force 
($\sim 1/d^3$), which agrees with the result based on 
Debye-H\"{u}ckel (Gaussian) approximation\cite{charge}.  
For the latter, a simple expression for the short-ranged force is derived, 
which shows that thermal fluctuations reduce 
its {\em range}, and which in the $T \rightarrow 0$ limit agrees with 
the known exponentially decaying result\cite{rb,zero}.  Furthermore, 
we argue that at zero temperature, there must also be a long-ranged 
force derived from the {\em quantum} fluctuations of the plasmons\cite{zero}, 
in addition to this zero-temperature exponentially decaying force.  
At very low temperatures in the quantum regime, we obtain a new length 
scale $\lambda_L$ (to be defined below), where the attraction scales 
like $d^{-2}$ when $d < \lambda_L$.

It should be pointed out that we have assumed a uniform charge 
distribution on the surface of the charged plates in our model for 
electrostatic attraction, mediated by the ``condensed'' counterions.  
This assumption of a uniform neutralizing background may not be a
good approximation to real experimental settings, since charges on 
macroion surfaces are discrete. For monovalent counterions, they
tend to bind to the charges on the surface and form dipolar molecules.
Therefore, the ground state for this system may not be a Wigner crystal, 
which relies on mutual repulsion among charges for its stability, 
and short-ranged effects are likely to be important.  However, for 
polyvalent counterions, a Wigner crystal is likely to form since 
each counterion does not bind to a particular charge on the surface, and
a uniform background may be more appropriate.  The detailed structure 
of the ground state as determined by short-ranged effects and valences will 
be the subject for another study.

Another point worth mentioning concerns 
the ordering of 2D solids which exhibit 
quasi-long-range-order (QLRO)\cite{lubensky}. It is 
well-known that a true long-range order is 
impossible for 2D systems with continuous symmetries.
For a 2D solid, which may be described by continuum elasticity theory with 
nonzero long-wavelength elastic constants, the Fourier components 
of the density function $n({\bf r})= 
\sum_{{\bf G}}\,n_{{\bf G}}({\bf r})\,e^{i\,{\bf G} 
\cdot {\bf r}}$ average (thermally) out to zero for a 
nonzero reciprocal lattice vector
${\bf G}$, {\em i.e.} $\langle n_{{\bf G}}({\bf r}) \rangle \,=\,
\langle e^{i\,{\bf G} \cdot {\bf u}({\bf r})}\rangle\, = 0,$ where
${\bf u}({\bf r})$ are the displacements of the particles from 
their equilibrium positions, while the correlation function decays 
algebraically to zero:
$\langle n_{{\bf G}}({\bf r})n^{*}_{{\bf G}}({\bf 0})\rangle\,\, 
\sim r^{-\eta_{{\bf G}}(T)}$ with 
$\eta_{{\bf G}}(T) = {k_B T G^2 (3 \mu + \lambda) \over 4\pi 
\mu (2\mu + \lambda)}$, where $\mu$ and $\lambda$ are 
Lam\'{e} elastic constants. This slow power-law decay of the
correlation function is very different from the exponential
decay one would expect in a liquid.  Hence the term QLRO.  
For a single 2D Wigner crystal, QLRO implies that 
the thermal average of the electrostatic potential at a 
distance $d$ above the plane is zero at any non-zero 
temperature, in contrast to a perfectly ordered lattice ($T=0$) 
where the electrostatic potential decays exponentially with $d$.  
This may lead to the conclusion that at finite temperatures
the short-ranged force between two coupled Wigner crystals 
should likewise be zero.  As we show below, this is not the case 
because the susceptibility, which measures the linear response 
of a 2D lattice to an external potential, nevertheless diverges at the 
reciprocal lattice vectors as in 3D solids\cite{diverge}.  

This paper is organized as follows.  In Sec. \ref{sec:eh}, we
derive an effective Hamiltonian which describes two interacting
planar Wigner crystals starting from the zero temperature 
ground state.  The total pressure is then decomposed into a long-ranged 
and a short-ranged component, which are evaluated in Sec. \ref{subsec:lr} and
\ref{subsec:sr}, respectively, and a detailed discussion of our
results is presented in Sec. \ref{subsec:results}.  
In Sec. \ref{sec:quantum}, we present an argument for a 
long-ranged attractive force arising from the zero-point fluctuations 
at zero temperature.  In addition, we use the Bose-Einstein distribution 
to calculate the attractive long-ranged pressure in the quantum regimes.

\section{Effective Hamiltonian and Pressure}
\label{sec:eh}

We start with the Hamiltonian for two interacting Wigner crystals:
${\cal H} = {\cal H}_0 + {\cal H}_{int}$.  Here, ${\cal H}_0$ is 
the elastic Hamiltonian for two isolated Wigner crystals\cite{2dwigner}\bb
\beta {\cal H}_0 = 
{1 \over 2}\, \sum_{i} \,\int {d^2 {\bf q} \over (2 \pi)^2} \, 
\Pi_{\alpha \beta}({\bf q})
\,\,u_{\alpha}^{(i)}({\bf q})\,u_{\beta}^{(i)}(-{\bf q}),
\label{isolated}\en
where $\beta^{-1} = k_B T$, ${\bf u}^{(i)}({\bf q}) $ is the Fourier 
transform of the in-plane displacement field of the charges in the
$i$th layer ($i=A$ or $B$), 
$\Pi_{\alpha \beta}({\bf q}) = \left [\, { 2 \pi l_B n^2 \over q }
\,P^{L}_{\alpha \beta} + \mu\,P^{T}_{\alpha \beta}\,\right]q^2$ 
is the dynamical matrix, $\mu \approx 
0.245\,n^{3/2}\,{l_B}$ 
is the shear modulus\cite{shear} in units of $k_B T$,
$P^{L}_{\alpha \beta} = q_{\alpha}q_{\beta}/q^2$ and
$P^{T}_{\alpha \beta} = \delta_{\alpha \beta} - 
q_{\alpha}q_{\beta}/q^2$ are longitudinal and transverse projection
operator, respectively.  Here, Greek indices indicate Cartesian components.
${\cal H}_{int}$ is the 
electrostatic interaction between the two layers:\bb
\beta  {\cal H}_{int} = l_B\,\int d^{2}{\bf x}\,d^{2}{\bf x}' 
\,{[\,\rho_A({\bf x}) - n\,]\,[\,\rho_B({\bf x}') - n\,] 
\over \sqrt{ ({\bf x} -{\bf x}')^2 + d^2\,}},
\label{inter}
\en
where $\rho_{i}({\bf x})$ is the number density of charges in the $i$th layer.
In order to capture the long-wavelength coupling as well as discrete 
lattice effects which are essential for our discussions on the 
short-ranged force,
we employ a method, similar to that in Ref.\cite{lattice},
which allows us to derive an effective Hamiltonian that
is valid in the elastic regime where the density fluctuations
are slowly varying in space, {\it i.e.} 
$\nabla \cdot {\bf u}^{(i)}({\bf x}) \ll\,1$,
but $|{\bf u}^{A}({\bf x}) - {\bf u}^{B}({\bf x})|$ 
need not be small compared to the lattice constant $a$.

Let us introduce a slowly varying field for each layer:\bb 
\phi^{(i)}_\alpha({\bf x})= x_\alpha 
- u_{\alpha}^{(i)}[\vec{\phi}^{(i)}({\bf x})],
\label{slow}
\en 
where the displacement field ${\bf u}^{(i)}({\bf x})$ is defined 
in such a way that it has no Fourier components outside 
of the Brillouin Zone (BZ).  Then, the density $\rho_i({\bf x})$ can 
be written as:\bb
\rho_i({\bf x}) = \sum_{l} \delta^{2}({\bf R}_l - \vec{\phi}^{(i)}({\bf x}))\,
\det[\partial_{\alpha}\,\phi^{(i)}_{\beta}({\bf x})],
\en
where ${\bf R}_l$ are the equilibrium positions of the charges, {\em i.e.}
the underlying lattice sites.  Using the Fourier representation of 
the delta function and solving $\phi^{(i)}_\alpha({\bf x})$ iteratively 
in terms of the displacement field, we obtain a decomposition 
of the density for the $i$th layer into a slowly
and a rapidly spatially varying pieces:\bb
\rho_i({\bf x}) - n \cong  -\,n\,\nabla \cdot {\bf u}^{(i)}({\bf x}) + 
\,\sum_{{\bf G} \neq 0}\,n\,e^{i\,{\bf G} \cdot [\,{\bf x} + 
{\bf u}^{(i)}({\bf x})\,]},
\label{density}
\en
where ${\bf G}$ is a reciprocal lattice vector.  
Note that we have neglected terms that are products of the slowly and 
the rapidly varying terms.  Physically, the first term represents 
density fluctuations for wavelengths greater than the
lattice constant, and the second term represents the underlying 
lattice, modified by thermal fluctuations.  Using the density 
decomposition (\ref{density}), ${\cal H}_{int}$ may be written 
as\begin{eqnarray}
\beta{\cal H}_{int} &=& \int\,{d^2{\bf q} \over (2 \pi)^2}\,
{ 2 \pi l_B   \over q }e^{ -q d}\,
\int d^2{\bf x}\,\int d^2{\bf x}'\,e^{i {\bf q}\cdot ( {\bf x} - {\bf x}')}\,
\left (\,n\,\nabla \cdot {\bf u}^{A}({\bf x}) - 
\,\sum_{{\bf G} \neq 0}\,n\,e^{i\,{\bf G} \cdot [\,{\bf x} + 
{\bf u}^{A}({\bf x})\,]} \right )\nonumber\\ 
&& \phantom{\,\sum_{{\bf G} \neq 0}\,n\,e^{i\,{\bf G} \cdot [\,{\bf x} + 
{\bf u}^{A}({\bf x})\,]}+{\bf u}^{A}({\bf x})}\times\left ( 
\,n\,\nabla \cdot {\bf u}^{B}({\bf x}') -
\,\sum_{{\bf G}' \neq 0}\,n\,e^{i\,{\bf G}' \cdot [\,{\bf x}' + \,{\bf c}\, + 
{\bf u}^{B}({\bf x}')\,]}\right ), 
\end{eqnarray}
where ${\bf c}$ is the relative displacement vector between 
two lattices of the different plane and we have used the fact that
${ 1 \over \sqrt{x^2 + d^2}} = 
\int {d^2{\bf q} \over (2 \pi)^2}\,e^{i {\bf q}\cdot{\bf x}}\,
{  2 \pi \over q }\,e^{ -q d}.$
Again neglecting the products of slowly and rapidly varying terms, 
which give vanishingly small contributions when integrating over all space, 
${\cal H}_{int}$ separates into two pieces: a long-wavelength term\bb
\beta  {\cal H}^{L}_{int} = \int\,{d^2{\bf q} \over (2 \pi)^2}\, 
{ 2 \pi l_B  n^2 \over q }e^{ -q d}\,q_{\alpha} q_{\beta}
\,u_{\alpha}^{A}({\bf q})\,u_{\beta}^{B}(-{\bf q}), 
\label{longh}
\en
and a short-wavelength term
\begin{eqnarray}
\beta {\cal H}^{S}_{int} = + \sum_{{\bf G} \neq 0}\,\sum_{{\bf G}' \neq 0}\,
\int\,{d^2{\bf q} \over (2 \pi)^2} &\,& { 2 \pi l_B  n^2 \over q }e^{ -q d}\, 
\nonumber \\
\times &\,&\int d^2{\bf x}\,\int d^2{\bf x}'\,
e^{i\,{\bf q} \cdot ({\bf x}- {\bf x}')}\,
\,e^{i\,{\bf G} \cdot [\,{\bf x} + {\bf u}^{A}({\bf x})\,]}\,
\,e^{i\,{\bf G}' \cdot [\,{\bf x}' +\,{\bf c}\,+ {\bf u}^{B}({\bf x}')\,]}. 
\label{ints1}
\end{eqnarray}
In order to obtain a tractable analytical treatment, 
we approximate this expression 
by splitting the sum over ${\bf G}'$ into two parts.  The dominant part, 
with ${\bf G}' = -{\bf G}$ is\bb
\beta {\cal H}^{S}_{int} = - \sum_{{\bf G} \neq 0}\,
\int\,{d^2{\bf q} \over (2 \pi)^2}\,{ 2 \pi l_B  n^2 \over q }e^{ -q\,d}\, 
\int d^2{\bf x}\,\int d^2{\bf x}'\,
e^{i\,({\bf q} + {\bf G}) \cdot ({\bf x}- {\bf x}')}\,
\,e^{i\,{\bf G} \cdot [\,{\bf u}^{A}({\bf x})\, - {\bf u}^{B}({\bf x}')]},
\label{ints2}
\en
where we have used $e^{i {\bf G}\cdot{\bf c}} = -1$.
The second part (those terms with ${\bf G}' \neq  -{\bf G}$) 
contains extra phase factors which tend to average to zero 
in the elastic limit.  As a first approximation, we neglect such terms. 
Finally, Eq. (\ref{ints2}) can be systematically expanded using a gradient 
expansion:\bb
\beta {\cal H}^{S}_{int} = 
- \sum_{{\bf G} \neq 0} \Delta_{\bf G}(d) \int d^{2}{\bf x}\,
\cos \left \{ {\bf G} \cdot \left [{\bf u}^{A}({\bf x}) - 
{\bf u}^{B}({\bf x}) \right ] \right \} + 
O(\partial_{\alpha} u_{\beta}^{(i)} \partial_{\gamma} u_{\tau}^{(j)}),
\label{cosine}
\en
where $\Delta_{\bf G}(d) =  { 4 \pi l_B n^2 \over G } e^{ -G d}.$
Putting Equations (\ref{isolated}), (\ref{longh}), and 
(\ref{cosine}) together, we obtain an effective Hamiltonian
for the coupled planar Wigner crystals:
\begin{eqnarray}
\beta {\cal H}_{e} =  \beta {\cal H}_{0} 
+  \int \,{d^2{\bf q} \over (2 \pi)^2}\,
{2 \pi l_B n^2 \over q }\,& e^{-qd}&\,q_{\alpha} q_{\beta}
\,u_{\alpha}^{A}({\bf q})\,u_{\beta}^{B}(-{\bf q})
\nonumber
\\
& -  & \sum_{{\bf G} \neq 0} \Delta_{\bf G}(d) \int d^{2}{\bf x}\,
\cos \left \{ {\bf G} \cdot \left [{\bf u}^{A}({\bf x}) - 
{\bf u}^{B}({\bf x})\right ] \right \}.
\label{effective}
\end{eqnarray}
The second term in Eq. (\ref{effective}) comes 
from the long-wavelength couplings while the third term 
reflects the periodicity of the underlying lattice structure.
This particular structure in the effective Hamiltonian, 
as will be demonstrated below, leads to a total force which is  
comprised of two pieces -- an exponentially decaying 
(short-ranged) force and a long-ranged power-law force:\bb
\Pi(d) = - {1 \over A_0}\,\left \langle { \partial \,{\cal H}_{int} 
\over { \partial d}} \right \rangle_{{\cal H}_e} = 
- {1 \over A_0}\left \langle { \partial \,{\cal H}^{S}_{int} 
\over { \partial d}} \right \rangle_{{\cal H}_e} -
{1 \over A_0}\,\left \langle { \partial \,{\cal H}^{L}_{int} 
\over { \partial d}} \right \rangle_{{\cal H}_e}
= \Pi_{SR}(d)+ \Pi_{LR}(d),
\label{force}
\en
where $A_0$ is the area of the plane.  It is important to emphasize 
that both forces are present 
simultaneously, although each force dominates at a 
different spatial scale -- the long-ranged force dominates
at large separations while the short-ranged force at small separations.

To calculate various expectation values in Eq. (\ref{force}), it is 
convenient to transform the displacement fields 
into in-phase and out-of-phase displacement fields by 
${\bf u}^{+}({\bf x}) = {\bf u}^{A}({\bf x}) + {\bf u}^{B}({\bf x})$ and 
${\bf u}^{-}({\bf x}) = {\bf u}^{A}({\bf x}) - {\bf u}^{B}({\bf x})$,
respectively, so that the effective Hamiltonian (\ref{effective}) 
separates into two independent parts: 
${\cal H}_e = {\cal H}_{+} + {\cal H}_{-}$ with\bb
\beta {\cal H}_{+} = {1 \over 2}\, \int {d^2 {\bf q} \over (2 \pi)^2} \, 
\Pi^{+}_{\alpha \beta}({\bf q})
\,\,u^{+}_{\alpha}({\bf q})\,u^{+}_{\beta}(-{\bf q}),
\label{har1}
\en
and\bb
\beta {\cal H}_{-} = {1 \over 2}\, \int {d^2 {\bf q} \over (2 \pi)^2} \, 
\Pi^{-}_{\alpha \beta}({\bf q})
\,\,u^{-}_{\alpha}({\bf q})\,u^{-}_{\beta}(-{\bf q})
- \sum_{{\bf G} \neq 0} \Delta_{\bf G}(d) \int d^{2}{\bf x}\,
\cos[{\bf G} \cdot {\bf u}^{-}({\bf x})],
\label{relative}
\en
where $\Pi^{\pm}_{\alpha \beta}({\bf q}) = {1 \over 2}\, 
\left [ {2\pi l_B n^2 \over q}( 1 \pm e^{-qd})\,P^{L}_{\alpha \beta} + 
\mu \,P^{T}_{\alpha \beta} \right ]\,q^2$.  Furthermore, 
at low temperature, where $\mid {\bf u}^{-}({\bf x}) \mid$ is small compared 
to the lattice constant $a$, the cosine term in Eq. (\ref{relative}) 
can be expanded up to second order in $|{\bf u}^{-}({\bf x})|$
to obtain the ``mass'' terms. Within a harmonic approximation ${\cal H}_{-}$,
up to an additive constant, may be written as
\bb
\beta {\cal H}_{-} \simeq  {1 \over 2}\, \int {d^2 {\bf q} \over (2 \pi)^2} \, 
\Pi^{-}_{\alpha \beta}({\bf q})
\,\,u^{-}_{\alpha}({\bf q})\,u^{-}_{\beta}(-{\bf q})
+ {1\over 2}\,\int {d^2 {\bf q} \over (2 \pi)^2}\,
\left [\,m_{L}^2 P^{L}_{\alpha \beta}
+m_{T}^2\,P^{T}_{\alpha \beta}\,\right ]
u^{-}_{\alpha}({\bf q})\,u^{-}_{\beta}(-{\bf q}).
\en
Here, $m_{L,T}^2 = 4 \pi l_B n^2 \,
\sum_{{\bf G} \neq 0} \, G\,e^{ -G d}= 4 \pi l_B n^2 \Delta_0(d)$. 
Note that the mass terms 
vanish exponentially with $d$ as also found in Ref.\cite{Bi}.
The fact that the transverse $m_{T}$ and longitudinal ``mass'' 
$m_{L}$ are degenerate is related to the underlying triangular structure of the
lattices\cite{Bi}.  These ``masses'' are associated with the finite 
energy required to uniformly shear the two Wigner crystals, 
and thus give rise to a gap in the dispersion relations 
of the out-of-phase modes.  In the next two subsections, we 
derive expressions for the long-ranged and the short-ranged 
pressure as given in Eq. (\ref{force}) within the 
harmonic approximation.

\subsection{Long-ranged Pressure}
\label{subsec:lr}

The long-ranged power-law force comes from the correlated 
long-wavelength density fluctuations (the plasmon modes).  
The shear modes do not contribute to this interaction since
$\partial_{\alpha}\,P^{T}_{\alpha \beta}\,u^{(i)}_{\beta}({\bf x}) = 0.$
Using Eqs. (\ref{longh}) and (\ref{force}), we obtain an expression 
for the long-ranged force:\bb
\beta \Pi_{LR}(d) = { 2 \pi l_B \over A_0} \,\int 
{d^2 {\bf q} \over (2 \pi)^2} \, e^{-qd} \, \langle\, 
\delta\rho_A({\bf q})\,\delta\rho_B(-\,{\bf q})\,\rangle,
\label{lf}
\en
where $\delta \rho_i({\bf x}) =  -\,n\,\nabla \cdot {\bf u}^{(i)}({\bf x})$ 
is the long-wavelength density fluctuation to the lowest order.
Making use of the equipartition theorem to evaluate\begin{eqnarray}
\langle\, 
\delta\rho_A({\bf q})\,\delta\rho_B(-\,{\bf q})\,\rangle &\propto&
\int D{\bf u}^{\pm}({\bf q})\,
\delta \rho_A({\bf q})\,\delta\rho_B(-\,{\bf q})\,
e^{-\beta[\cal{H}_{+} +  \cal{H}_{-}]} \nonumber\\
&=& - {A_0}\,{n^2\,q^2 \over 4  } 
\left [ {1 \over \pi l_B n^2 q\,(1 - e^{-qd}) + 
m_L^2} - {1 \over \pi l_B n^2 q\,(1 + e^{-qd})} \right ], 
\end{eqnarray}
and substituting this into Eq. (\ref{lf}), 
we have the result\bb
\Pi_{LR}(d) = -\,{k_B T \over d^3 }\,\alpha(\Delta_0\,d), 
\en
where\bb
\alpha(x) \cong \,{\zeta(3) \over 8 \pi} + {x \over \pi}
\,\left [\, \mbox{Ci}\,(\,2\sqrt{x}\,)\,\cos(\,2\sqrt{x}\,) + 
\mbox{Si}\,(\,2\sqrt{x}\,)\,\sin(\,2\sqrt{x}\,) \,\right ],
\label{alpha}
\en
$\zeta$ is the Riemann zeta function, and Ci$(x)$ and Si$(x)$
are the cosine and sine integral functions\cite{handbook}, respectively.  
In the large distance limit, the second term 
in Eq. (\ref{alpha}) is exponentially suppressed and can 
be neglected, yielding $\alpha  = {\zeta(3) \over 8 \pi}$.  
Therefore, for large $d$ we have\bb
\Pi_{LR}(d) = -\,{\zeta(3) \over 8 \pi}\,
{k_B T \over d^{\,3}}.
\label{lr2}
\en
This is the well-known result from the Debye-H\"{u}ckel 
approximation\cite{charge}.  Note also that the amplitude 
${\zeta(3) \over 8 \pi} \cong 0.048$ is universal for
this interaction, induced by the long wavelength fluctuations\cite{kardar}.

\subsection{Short-ranged Pressure}
\label{subsec:sr}

The short-ranged force which decays exponentially owes its existence to 
the ``structural'' correlations.  It survives even at non-zero temperature, 
in contrast to the conclusion drawn from a single 2D Wigner 
crystal, as discussed in the Introduction.  
However, we expect on physical grounds 
the short-ranged force to be weakened by thermal fluctuations.  
To compute its temperature dependence explicitly, we start with 
the expression for this 
force derived from Eqs. (\ref{ints2}) and (\ref{force}):\bb
\beta \Pi_{SR}(d) =  -\,2 \pi l_B n^2 \sum_{{\bf G} \neq 0}\,
e^{-\,{ {G}^2 \over 2} \langle |\,{\bf u}^{-}(0)|^2 \rangle \,}
\,f_{\bf G}(d),
\label{short2}
\en
where $f_{\bf G}(d) = \int\,{d^2{\bf q} \over (2 \pi)^2}\,
{\cal S}({\bf q} - {\bf G})\,e^{ -q d}\,$, 
${\cal S}({\bf q} - {\bf G}) = \int\,d^2{\bf r}
\,e^{i\,({\bf q}- {\bf G}) \cdot {\bf r}}\,
e^{-{ {G}^2 \over 8}\,[ \,B^{+}({\bf r})\, - \, B^{-}({\bf r})]\,}$, 
and $B^{\pm}({\bf r}) = \langle\,[ {\bf u}^{\pm}({\bf r}) 
- {\bf u}^{\pm}({\bf 0}) ]^2\,\rangle$.  Note that Eq. (\ref{short2}) 
is exact, provided all the averages are evaluated exactly.
For a system of coupled perfect Wigner crystals 
at zero temperature, $f_{\bf G}(d) = e^{-Gd}$.  
At finite temperature, but below the melting temperature $T_m$, we note that 
$B^{\pm}(r)$ varies very slowly in space, so that 
$f_{\bf G}(d)$ can be approximated by its zero temperature value: 
$f_{\bf G}(d) \simeq e^{-Gd}$.  Hence, we obtain\bb
\beta \Pi_{SR}(d) \cong -\,2 \pi l_B n^2 \sum_{{\bf G} \neq 0}\,e^{-Gd}
\,\left \langle e^{i\, {\bf G} 
\cdot [{\bf u}^{A}({\bf 0})- {\bf u}^{B}({\bf 0})]} \right 
\rangle_{{\cal H}_e}.
\label{short1}
\en
The thermal average of the displacement fields 
in Eq. (\ref{short1}) resembles a ``Debye-Waller'' factor
and indicate the degree to which the short-ranged force is 
depressed by thermal fluctuations from 
its zero temperature maximum value. Because of the cosine
term present in Eq. (\ref{effective}), this ``Debye-Waller'' 
factor is in general not zero, unlike the case of a single 
2D Wigner crystal.  However, if the system has melted into a Coulomb fluid,
this cosine term, which comes from the lattice structure, 
would have to be modified.

The required expectation value in Eq. (\ref{short1})
only involves $\cal{H}_{-}$.  Within the harmonic approximation, 
the mean-square out-of-phase displacement field can be evaluated\bb
\langle |{\bf u}^{-}({\bf x})|^2 \rangle
\cong  {\lambda_D \over 2 \pi n d} \ln \left [ { d \over
4 \Delta_0(d)a^2} \right ] + {1 \over 2 \pi \mu}\, 
\ln \left [ {\mu \over 8\pi l_B n^2 \Delta_0(d)a^2} \right ]
\cong {G_0 d  \over 2 \pi }\,\left [ { \lambda_D \over n d} + {1 \over \mu}
\right ], 
\label{hsr}
\en
where $\lambda_D = 1/( 2 \pi l_B n)$,
$a$ is the lattice constant, $\mu \approx 0.245\,n^{3/2}\,{l_B}$ 
is the shear modulus of an isolated Wigner crystal in units of $k_B T$,
and in the last line, we have approximated 
$\,\Delta_0(d)\,$ by the first nonzero 
reciprocal lattice vector contribution: $\Delta_0(d) \approx G_0\,e^{-G_0 d}$.
Note also that the logarithmic dependence on the ``mass''
($= 4 \pi n^2 l_B \Delta_0(d)$) is a characteristic of 2D solids.  
Inserting Eq. (\ref{hsr}) into Eq. (\ref{short1}), we obtain 
an expression for the short-ranged pressure at finite temperatures\bb
\beta \Pi_{SR}(d) \simeq  -\,2 \pi l_B n^2 \,
e^{- ( 1 + \,\xi/2 )\,G_0 d}.
\label{sh}
\en
Here, the parameter $\xi$ defined by\bb
\xi = { G_0^2 \over 2\pi }\,
\left (\,  { \lambda_D \over n d} + {1 \over \mu} \, \right ),
\en
characterizes the relative strengths of thermal fluctuations
and the electrostatic energy of a Wigner crystal, {\em i.e.} 
$\xi \sim {k_B T\,a \over e^2}$.  Thus, the sole effect of 
thermal fluctuations on the short-ranged
force is to reduce its  {\em range}: $G_0 \rightarrow G_0 
\left (\,1 + {\xi \over 2 } \, \right )$.

\subsection{Discussion of Results}
\label{subsec:results}

\begin{figure}
\epsfxsize=3.0in
\epsfysize=2.3in
\centerline{\epsfbox{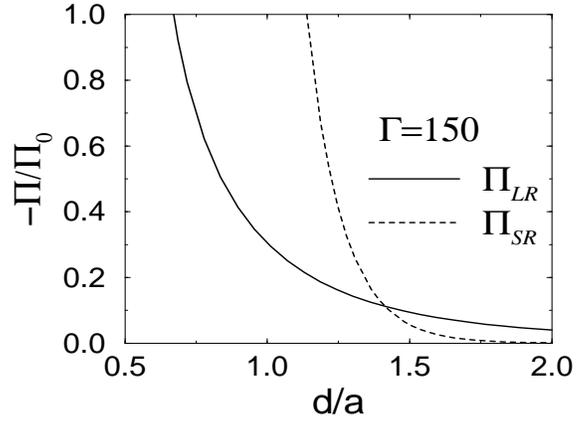}}
\centerline{\mbox{(a)}}
\epsfxsize=3.0in
\epsfysize=2.3in
\centerline{\epsfbox{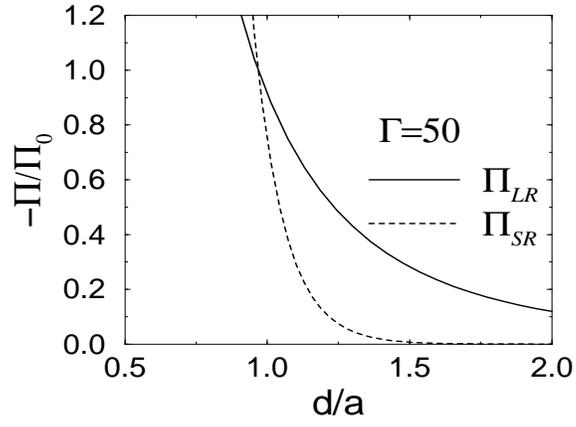}}
\centerline{\mbox{(b)}}
\vfil
\caption{Plots of $\Pi_{SR}$ and $\Pi_{LR}$ versus $d$ for $\Gamma = 150$ (a) 
and $50$ (b).  Observe that the crossover ($\Pi_{LR} \approx \Pi_{SR}$)
occurs at about $d \sim a$. $\Pi_{0} \equiv k_B T\,(l_B / a^4) 
\times 10^{-3}.$}
\label{prep}
\end{figure}

In summary, we have shown that the total pressure can be decomposed into a 
long-ranged $\Pi_{LR}$ and a short-ranged pressure $\Pi_{SR}$.  
Each force is computed at low temperatures, where the harmonic 
approximation is expected to be valid.  The result for the total force 
is\bb
\beta \Pi(d) \simeq  -\,2 \pi l_B n^2 \,
e^{- ( 1 + \,\xi/2 )\,G_0 d}  - \,{\alpha(\Delta_0 d ) \over d^3},
\label{total}
\en
where $\xi = { G_0^2 \over 2\pi }\,
\left [  { \lambda_D \over n d} + {1 \over \mu} \right ]$ and
$\alpha(\Delta_0 d) = {\zeta(3) \over 8 \pi}$ for large $d$.
In Fig. \ref{prep}, we have plotted $\Pi_{SR}$ 
and $\Pi_{LR}$ for two values of the coupling constant, 
$\Gamma \equiv {l_B \over a} = 150$ and $50$.  Not surprisingly, 
they show that $\Pi_{SR}$ dominates for small $d$, and 
$\Pi_{LR}$ for large $d$.  However, it is 
interesting to note that even for high values of $\Gamma$, 
$\Pi_{LR}$ dominates as soon as $d \sim a$.

\begin{figure}
\epsfxsize=6.0in
\epsfysize=4.6in
\centerline{\epsfbox{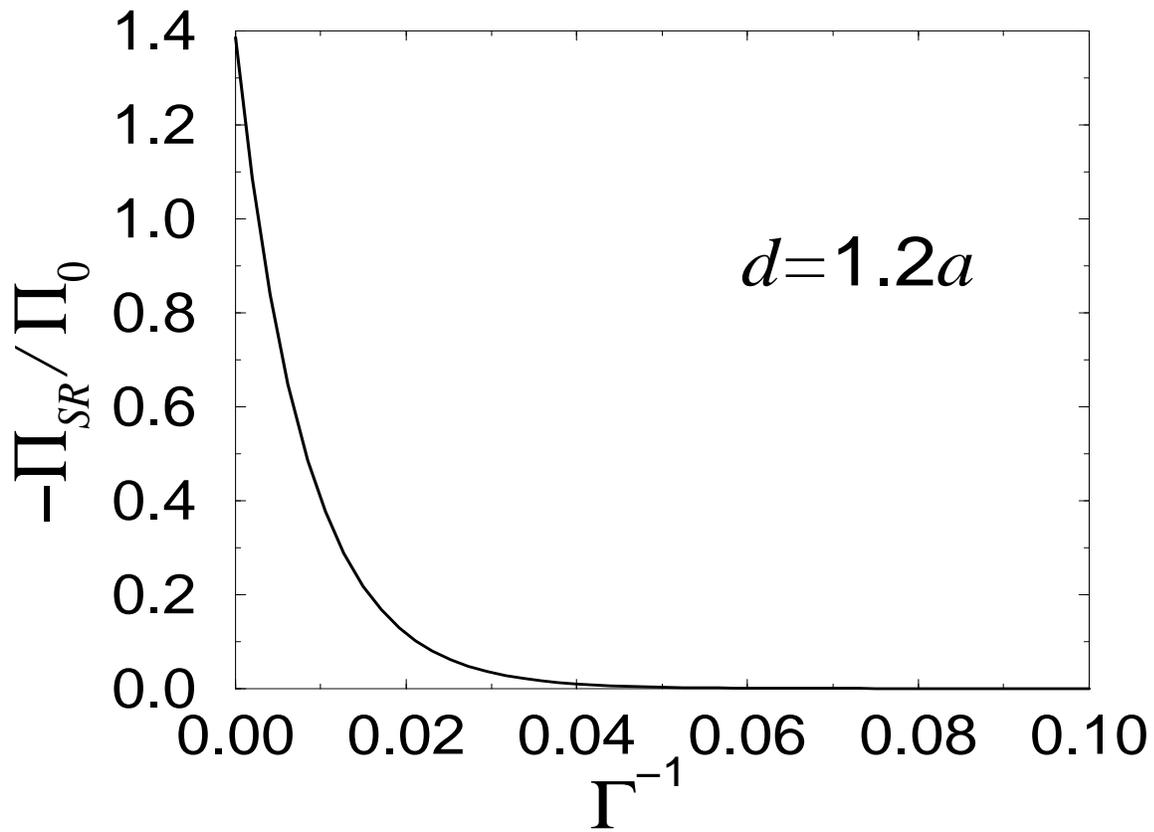}}
\vfil
\caption{A plot of $\Pi_{SR}$ as a function of  $\Gamma^{-1}$
at $d= 1.2\,a$, according to Eq. (\ref{sh}), which 
shows that $\Pi_{SR}$ exponentially decreases with increasing temperature.}
\label{srp}
\end{figure}

According to Eq. (\ref{sh}), the magnitude of $\Pi_{SR}$
tends to decrease exponentially with temperature, as illustrated in 
Fig. \ref{srp}.  This strong decrease with increasing temperature
is consistent with the Brownian dynamics simulations of Gr\o nbech-Jensen 
{\em et al.}\cite{attractionS}.  The shortening of its 
range may be attributed to the generic nature of strong 
fluctuations in 2D systems, and can also be understood 
by the following scaling argument.  Referring back to ${\cal H}_{-}$ 
in Eq. (\ref{relative}), one can show that the 
anomalous dimension of the operator 
$\cos[{\bf G}_0 \cdot {\bf u}^{-}({\bf x})]$ is $[$ Length $]^{-\xi}$ 
and correspondingly the dimension of $\Delta_{{\bf G}_0}(d)$ is
$[$ Length $]^{\xi - 2}$.  Since $\Delta_{{\bf G}_0}(d)$ is the only 
relevant length scale in ${\cal H}_{-}$, we must have 
$\left \langle e^{i\, {\bf G}_0 \cdot {\bf u}^{-}({\bf 0})} \right \rangle
\sim \Delta_{{\bf G}_0}^{\,\,\,{ \xi \over 2 -  \xi}}$\cite{sG}.  Therefore, 
the short-ranged pressure scales like\bb 
\Pi_{SR}(d) \sim  -\, \Delta_{{\bf G}_0}(d) \times 
\left \langle e^{i\, {\bf G}_0 \cdot {\bf u}^{-}({\bf 0})} \right \rangle
\sim -\,\Delta_{{\bf G}_0} \times 
\Delta_{{\bf G}_0}^{\,\,\,{ \xi \over 2 -  \xi}}
\sim - \,e^{- G_0 d \left ( {2 \over  2 - \xi} \right )}.
\label{scaling}
\en
In the low temperature limit ($\xi \ll 1$), we see that the range of
$\Pi_{SR}$ is $G_0 \left (\,1 + {\xi \over 2 } \, \right )$
as in Eq. (\ref{sh}).  This scaling argument also suggests that 
at higher temperatures thermal fluctuations may have interesting 
nonperturbative effects.  At zero temperature $\xi = 0$, so $\Pi_{SR}$ 
in Eq. (\ref{sh}) reproduces the known result of exponentially decaying 
attractive force\cite{rb,zero}.  

The long-ranged pressure for large $d$ in Eq. (\ref{lr2})
agrees exactly, including the prefactor, with the Debye-H\"{u}ckel 
approximation.  This is hardly surprising since the existence of 
long-wavelength plasmons (average density fluctuations) 
is independent of local structure, and they are present 
for solids and fluids alike.  Thus, the asymptotic long-ranged power-law 
force must manifest itself even after QLRO is lost via a 2D 
melting transition driven by dislocations\cite{dislocation}.   
Therefore, our formulation captures the essential physics 
of the attraction not only arising from the ground state 
``structural'' correlations, but also from 
the high temperature charge-fluctuations.

\section{Quantum Contributions to the Long-ranged Attraction}
\label{sec:quantum}

According to the {\em classical} calculations above, correlation effects 
give rise to a ``structural'' short-ranged and 
a long-ranged attractive force.  Recall that the long-ranged force 
vanishes as $T \rightarrow 0$, and that the short-ranged force is 
strongest at zero temperature but vanishes exponentially 
with distance.  This observation suggests that for sufficiently
large separations correlated attractions 
at finite temperatures are stronger than those arising from the 
perfectly correlated zero temperature ground
state.  However, we have pointed out in Ref. \cite{zero} that 
zero-point fluctuations of the {plasmons} lead to an attractive 
long-ranged interaction, which exhibits an unusual 
fractional-power-law decay ($\sim d^{-7/2}$), 
in contrast to the zero-temperature van der Waals 
interaction ($\sim d^{-4}$).
Hence, in the $T \rightarrow 0$ limit, this ``zero-point attraction''
dominates the short-ranged 
``structural'' force at large separations. Furthermore, 
we expect that quantum fluctuations persist at finite temperature,
and in this section, we compute their temperature dependence.

Within the harmonic approximation to the effective Hamiltonian, 
the dispersion relations for the plasmons can be 
readily obtained\cite{zero}:
\begin{eqnarray}
\omega_1^2(q) & = &
\frac{8 \pi e^2 n}{m \epsilon}\,\Delta_0(d) + \frac{2 \pi e^2 n}{m \epsilon}\,
q\,(\,1 - e^{-q d}\,); \\
\omega_2^2(q) & = &
\frac{2 \pi e^2 n}{m \epsilon}\,q\,(\,1 + e^{-q d}\,),
\end{eqnarray}
where $m$ is the mass of the charges and $\Delta_0(d) \sim e^{-G d}$ 
is proportional to the energy gap (the ``mass'' term) 
for the out-of-phase mode.  The plasmon modes are related 
to the correlated charge-density fluctuations
in the two layers.  At any finite temperature, 
the free energy of the low-lying plasmon excitations is given by\bb
{\cal F}(d)/A_0 = {\hbar \over 2} \sum_{i=1,2}\,
\int {d^2 {\bf q} \over (2 \pi)^2} \,\,\omega_i({\bf q})
+ k_B T \sum_{i=1,2}\,\int {d^2 {\bf q} \over (2 \pi)^2} \,\ln 
\left [1 - e^{- \beta \hbar \omega_i({\bf q})} \right ], 
\label{free}
\en
where $A_0$ is the area of the plane.  Since the energy gap 
$\Delta_0$ is exponentially damped for large distances, its 
contribution to the free energy may be neglected in the large distance limit, 
where the long-ranged force is expected to be dominant.

The first term in Eq. (\ref{free}) arising from the
zero-point fluctuations has been computed in Ref. \cite{zero} 
and gives the $d^{\,-7/2}$ power-law mentioned above.  An additional 
contribution to the pressure
at finite temperature arises from the second term in
Eq. (\ref{free}),
\begin{eqnarray}
\beta \Pi_{LR}(d) = - { \hbar \Lambda \over 4 \pi d^{\,7/2}}\,
\int_0^{\infty} dx\, x^{5/2}  \left 
\{ {1 \over \exp[\eta \sqrt{x(1-e^{-x})}] - 1}\,\right.
{e^{-x} \over \sqrt{ 1 - e^{-x}}} &\,&
\nonumber \\
-  {1 \over \exp[\eta \sqrt{x(1+e^{-x})}] - 1}&\,&
\left.  {e^{-x} \over \sqrt{ 1 + e^{-x}}}  
\vphantom {1 \over \exp[\eta \sqrt{x(1+e^{-x})}] - 1} \right\},
\label{pressure}
\end{eqnarray}
where $\Lambda = \sqrt{{2 \pi e^2 n \over m \epsilon}} $ and
$\eta = \beta \hbar \Lambda / \sqrt{d}$. We can evaluate this expression 
in two limits:

In the low-temperature limit $\,\eta \gg 1$, 
Eq. (\ref{pressure}) can be systematically expanded 
in powers of $\eta^{-1}$.  The lowest order term is given by 
$\Pi_{LR}(d) = -\overline{\alpha}\,{k_B T \over \lambda_L d^2}$,
where $\lambda_L \equiv {a_B}\,{ l_B \over 2\lambda_D},$
$a_B \equiv \epsilon \hbar^2/ (m e^2)$ is the effective Bohr radius, 
$\overline{\alpha} \equiv {1 \over 4 \pi} \int_0^{\infty}\,dx\,
{ x^2 \over e^{x} - 1} = \zeta(3)/(2\pi)$, 
and $\zeta$ is the Riemann zeta function.  
We observe that the low temperature condition $\eta > 1$ is 
equivalent to the short distance limit $d < \lambda_L$.

In the high temperature limit $\eta \ll 1$ or the large distance 
limit $d > \lambda_L$, we expand the exponential 
in the denominator of Eq. (\ref{pressure}) to obtain 
$\Pi_{LR}(d) = - \alpha\,{ k_B T \over  d^3},$ where $\alpha = 
\zeta(3)/(8\pi)$.  This result agrees with the classical calculation
in Sec. \ref{subsec:lr} as it should.  Therefore, we have 
the following regimes for correlated attraction from plasmon 
fluctuations at finite temperature\bb
\Pi_{LR}(d) \sim  \left \{  \begin{array}{ll}
                    -k_B T/d^3, & \mbox{ for $\lambda_L < d$,} \\
                    -k_B T/(\lambda_L d^2),  & \mbox{ for $\lambda_L > d.$}\\
                            \end{array} 
        \right.
\label{solid}
\en
We note that $\lambda_L$, in contrast to $\lambda_D$,
increases with decreasing temperature, indicating, 
as one might expect, that quantum fluctuations are important 
at low temperatures.  Furthermore, since $\Pi_{LR}(d) 
\rightarrow 0$ as $T \rightarrow 0$,
the attractive interaction as $T \rightarrow 0$
is governed by zero-point fluctuations as emphasized above.
In the strong Coulomb coupling limit 
$l_B/\lambda_D \sim 100$, we get $\lambda_L \sim 3\,$\AA$\,\,$for 
$\epsilon \sim 100$ and $a_B \sim 1/20\,$\AA.  Finally, 
it should be emphasized that quantum contributions to the 
long-ranged attraction are unlikely to be relevant for macroions.
Our motivation here stems from the desire to understand the 
charge-fluctuation-induced attraction between coupled layers in a 
complete picture.  However, our results may be relevant for electrons in 
bilayer semiconductor systems.  Indeed, there are recent theoretical 
efforts devoted to this subject \cite{qubi}.

\section{Discussion and Conclusion}
\label{sec:conclusion}

\begin{figure}
\epsfxsize=6.0in
\epsfysize=4.6in
\centerline{\epsfbox{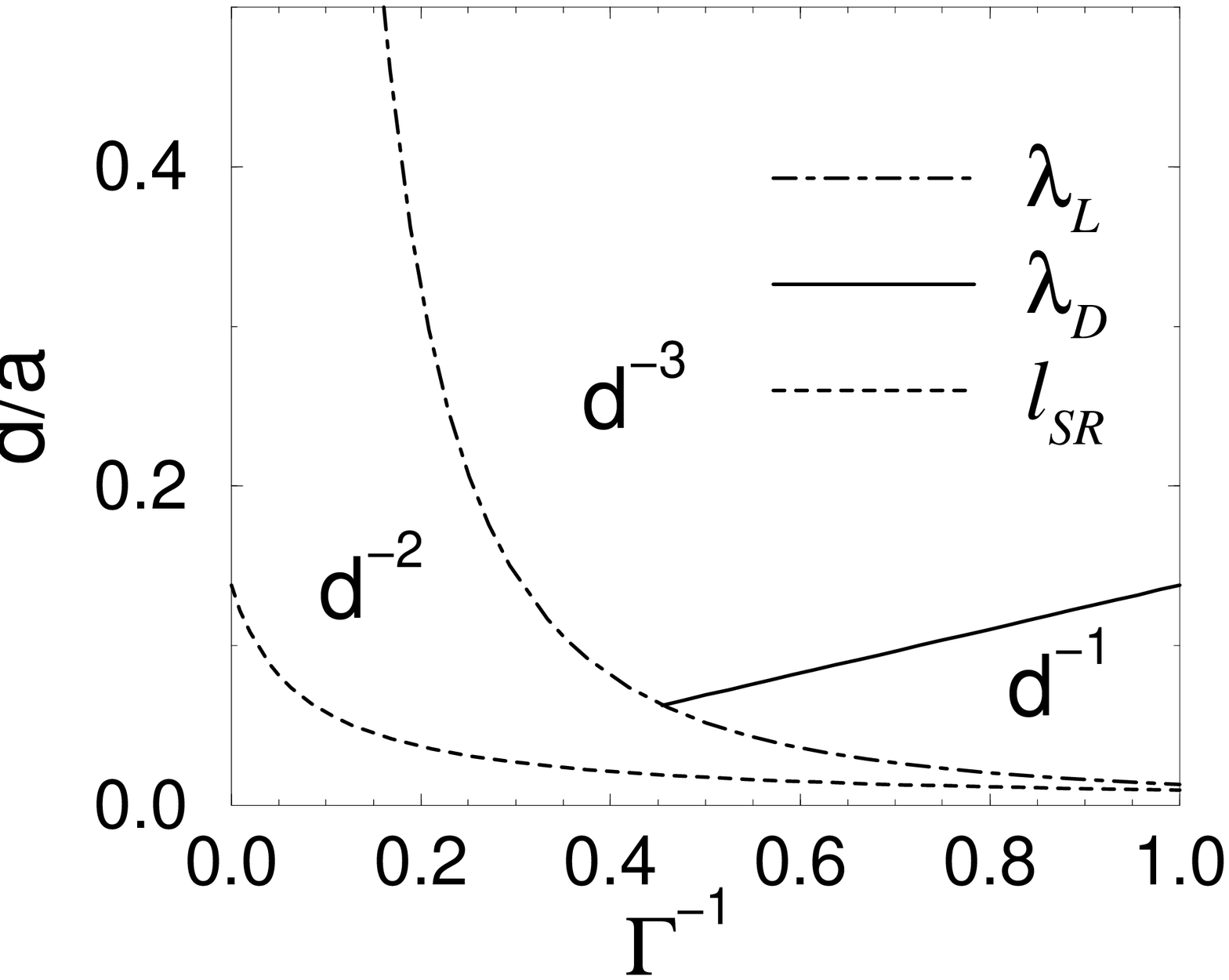}}
\vfil
\caption{A schematic phase diagram summarizing different 
charge-fluctuation-induced attraction regimes.  The characteristic decay 
length $l_{SR}$ of the short-ranged force is also shown.}
\label{phase}
\end{figure}

In this paper, we have studied analytically the electrostatic attraction
between two planar Wigner crystals in the strong Coulomb 
coupling limit.  We show that the total attractive pressure 
can be separated into a long-ranged and short-ranged component.  
The long-ranged pressure arises from {\em correlated fluctuations} and
the short-ranged pressure from the ground state {\em ``structural'' 
correlations}.  We also compute the very low temperature behavior
of the fluctuation-induced attraction, where long-wavelength plasmon 
excitation must be described by Bose-Einstein statistics.  The results 
are summarized in Fig. \ref{phase}, showing different regimes 
for the charge-fluctuation-induced long-ranged attraction, 
including the high temperature results in Ref. \cite{charge} and the 
characteristic decay length $l_{SR}$ for the short-ranged force.
For small $d$, the short-ranged force is always dominant, but the
decay length shrinks with increasing temperature.  The crossover from 
the short-ranged to long-ranged dominant regimes occurs about $d \sim a$.
Thus, for large $d \gg a$ only the long-ranged force is operative, which
crosses over from $d^{\,-7/2}$ at zero temperature to the finite temperature 
distance dependence of $d^{-2}$ if $d < \lambda_L$ and 
$d^{-3}$ if $d > \lambda_L$.  This provides a unified 
description to the electrostatic attraction 
between two coupled Wigner crystals.

In addition, our formulation may offer further insights 
into the nature of the counterion-mediated 
attraction at short distances.  As 
discussed in Sec. \ref{subsec:sr}, the reason that the short-ranged force 
in Eq. (\ref{short1}) does not vanish is because 
of the cosine term in ${\cal H}_{-}$, which represents the underlying
lattice structures, and our results indicate that the strength of the 
short-ranged force decreases exponentially with 
temperature.  However, at higher temperatures the expression for 
$\Pi_{SR}$ in Eq. (\ref{sh}) is no longer valid, since the harmonic 
approximation breaks down.  Indeed, the scaling argument leading to 
Eq. (\ref{scaling}) suggests that if the full cosine term is retained, 
$\Pi_{SR}$ may exhibit nonperturbative behaviors as $\xi \rightarrow 2^{-}$.

To discuss qualitatively what happens at higher temperatures, we assume 
that $\Delta_{\bf G}(d)$ is sufficiently small and the system of 
interacting Wigner crystals is below its melting temperature $T_m$.  
Then, the charges between the two layers may unlock via
a Kosterlitz-Thouless (KT) type of transition, determined by 
the relevancy of the cosine term in ${\cal H}_{-}$, at $\xi = 2$
\cite{fertig}. (An order of magnitude estimate for the coupling constant
is $\Gamma \sim 13$.)  In the locked phase, 
$\xi \ll 2$, the periodic symmetry in ${\cal H}_{-}$ is 
spontaneously broken, and the resulting 
state is well captured by the harmonic approximation.  
On the other hand, when $\xi > 2$ the fluctuations are 
so large that the ground state becomes nondegenerate (gapless), 
{\em i.e.} the layers are decoupled.  To compute $\Pi_{SR}$
in the unlocked phase, ${\cal H}^{S}_{int}$ given in Eq. (\ref{cosine})
can be treated as a perturbation in evaluating the ``Debye-Waller'' 
factor in Eq. (\ref{short1}).  To the lowest order, we obtain\bb
\Pi_{SR}(d) \simeq - {k_B T \over \lambda_D^2 a} 
\,\left (\,{ \xi - 1 \over \xi - 2}\, \right )\,{e^{ -2 G_0 d}}\,. 
\label{srh}
\en
We first note that this expression diverges as $\xi \rightarrow 2^{+}$, 
indicating the breakdown of the perturbation theory as the 
temperature is lowered.   
Furthermore, in contrast to Eq. (\ref{sh}),
the range of $\Pi_{SR}$ remains constant and the amplitude acquires 
a temperature dependence of $\sim 1/T$ (for large $\xi \gg 2$), 
reminiscent of a high temperature expansion.

However, the above picture may be modified if the charges have
melted into a Coulomb fluid via a dislocation-mediated 
melting transition\cite{dislocation} before $\xi \rightarrow 2^{-}$.  
If this is the case, further analysis is necessary to obtain a 
more complete picture of the high temperature phase.  Although 
the spatial correlations in a system of coupled 2D Coulomb fluids are 
expected to be somewhat different from 2D Wigner crystals, 
the solid phase results above suggest a qualitative {\em lower} limit of 
$\Gamma \sim 13$ at which $\Pi_{SR}$ crosses over from low temperature 
to high temperature behavior.  It may be of interest to 
note that in Ref. \cite{roland}, an estimate for the {\em upper} 
limit of $\Gamma$ at which the Poisson-Boltzmann equation breaks 
down is of the order of $\Gamma \sim 3$.  To describe the melting of coupled 
2D Wigner crystals, excitations of dislocations must be introduced into 
the effective Hamiltonian Eq. (\ref{effective}) similar 
to what is done in Ref. \cite{sliding}.  These considerations 
may help to establish an analytical theory of the attraction arising from 
counterion correlations, not captured by the Poisson-Boltzmann theory.  
The present formulation is a first step in that direction.

\section{acknowledgment}
\label{acknowledgement}

We would like to thank Ramin Golestanian,
T. C. Lubensky, A.W.W. Ludwig, and S. Safran for stimulating 
and helpful discussions.  AL and PP acknowledge support from NSF 
grants MRL-DMR-9632716, DMR-9624091, and DMR-9708646. DL acknowledges 
support from Israel Science Foundation grant 211/97. HF acknowledges 
support from NSF Grant No. DMR-9870681.

\pagebreak

\newpage

\newpage

\newpage

\newpage

\begin{thebibliography}{99}

\bibitem{nature}
J. N. Israelachvili, {\em Intermolecular and Surface Forces.}
(Academic Press Inc., San Diego, 1992).

\bibitem{attractionE}
V. A. Bloomfield, Biopolymers {\bf 31}, 1471 (1991); R. Podgornik, D. Rau,
and V. A. Parsegian, Biophys. J. {\bf 66}, 962 (1994); A. E. Larsen and
D. G. Grier, Nature {\bf 385}, 230 (1997).

\bibitem{attractionS}
L. Guldbrand, B. J\"{o}nsson, H. Wennerstr\"{o}m, and P. Linse, 
J. Chem. Phys. {\bf 80}, 2221 (1984); S. Marcelja, Biophys. J. 
{\bf 61}, 1117 (1992); M.J. Stevens and K. Kremer, J. Chem. Phys. 
{\bf 103}, 1669 (1995); N. Gr\o nbech-Jensen, R. J. Mashl, 
R.F. Bruinsma, and W.M. Gelbart, Phys. Rev. Lett. {\bf 78}, 2477 (1997); 
E. Allahyarov, I. D'Amico, and H. L\"{o}wen, Phys. Rev. Lett. {\bf 81},
1334 (1998); N. Gr\o nbech-Jensen, K.M. Beardmore, and P. Pincus,
Physica A, {\bf 261}, 74 (1998).

\bibitem{thm}
J.C. Neu, Phys. Rev. Lett. {\bf 82}, 1072 (1999); J.E. Sader
and D.Y. Chan, J. Colloid Interface Sci. {\bf 213}, 268 (1999).

\bibitem{manning}
G. S. Manning, J. Chem. Phys. {\bf 51}, 924 (1969); S. Alexander, P. M.
Chaikin, P. Grant, G. J. Morales, P. Pincus, and D. Hone, J. Chem. Phys. 
{\bf 80}, 5776 (1984).


\bibitem{roland}
R. Netz and H. Orland, Eur. Phys. J. E {\bf 1}, 203 (2000).

\bibitem{charge}
Phil Attard, Roland Kjellander, and D. John Mitchell, 
Chem. Phys. Lett. {\bf 139}, 219 (1987); B.-Y. Ha and A. J. Liu, 
Phys. Rev. Lett. {\bf 79}, 1289 (1997); P. Pincus and 
S.A. Safran, Europhys. Lett. {\bf 42}, 103 (1998); D. B. 
Lukatsky and S. A. Safran, Phys. Rev. E {\bf 60}, 5848 (1999).

\bibitem{rb}
I. Rouzina and V. A. Bloomfield, J. Phys. Chem. {\bf 100}, 9977 (1996);
B. I. Shklovskii, Phys. Rev. Lett. {\bf 82}, 3268 (1999); J. Arenzon,
J. F. Stilck, and Y. Levin, Eur. Phys. J. B. {\bf 12}, 79 (1999).

\bibitem{zero}
A.W.C. Lau, Dov Levine, and P. Pincus, Phys. Rev. Lett. {\bf 84}, 4116 (2000).

\bibitem{lubensky}
P.M. Chaikin and T.C. Lubensky, {\em Principles of Condensed Matter 
Physics} (Cambridge Univ. Press, NY, 1995).

\bibitem{diverge}
Y. Imry and L. Gunther, Phys. Rev. B {\bf 3}, 3939 (1971).

\bibitem{2dwigner}
For a discussion of the elasticity theory for 2D Wigner crystals, 
see Daniel S. Fisher, B.I. Halperin, 
and R. Morf, Phys. Rev. B. {\bf 20}, 4692 (1979); 
Danial S. Fisher, Phys. Rev. B {\bf 26}, 5009 (1982).


\bibitem{shear}
L. Bonsall and A. A. Maradudin, Phys. Rev. B {\bf 15}, 1959 (1977).

\bibitem{lattice}
F.D.M. Haldane, Phys. Rev. Lett. {\bf 66}, 2270 (1991);
Thierry Giamarchi and Pierre Le Doussal, Phys. Rev. Lett. {\bf 72}, 1530
(1994); Phys. Rev. B {\bf 52}, 1242 (1995).

\bibitem{Bi}
G. Goldoni and F. M. Peeters, Phys. Rev. B {\bf 53}, 4591 (1996);
K. Esfarjani and Y. Kawazoe, J. Phys.: Condens. Matter {\bf 7}, 7217 (1995);
V. I. Falko, Phys. Rev. B {\bf 49}, 7774 (1994).


\bibitem{handbook}
{\em Handbook of Mathematical Functions}, edited by M. Abramowitz, 
I.A. Stegun (Dover Publications, New York, 1972).

\bibitem{kardar}
Mehran Kardar and Ramin Golestanian, Rev. Mod. Phys. {\bf 71}, 
1233 (1999).

\bibitem{sG}
A. B. Zamolodchikov and S. Lukyanov, Nucl. Phys. B {\bf 493}, 571 (1997);
V. Fateev, S. Lukyanov, A. B. Zamolodchikov, and Al. B. Zamolodchikov,
Phys. Lett. B {\bf 406}, 83 (1997).


\bibitem{dislocation}
For a review of dislocation-mediated melting transition
for 2D solids, see D. R. Nelson, in {\em Phase Transitions and
Critical Phenomena}, eds. C. Domb and J. L. Lebowitz 
(Academic Press, New York, 1983).

\bibitem{qubi}
B.E. Sernelius and P. Bj\"{o}rk, Phys. Rev. B {\bf 57}, 6592 (1998);
J.F. Dobson and J. Wang, Phys. Rev. Lett. {\bf 82}, 2123 (1999);
M. Bostr\"{o}m and Bo E. Sernelius, Phys. Rev. B {\bf 61}, 2204 (2000).


\bibitem{fertig}
H. A. Fertig, Phys. Rev. Lett. {\bf 82}, 3693 (1999).


\bibitem{sliding}
C. S. O'Hern, T. C. Lubensky, and J. Toner, Phys. Rev. Lett. {\bf 83}, 
2745 (1999).

\end{thebibliography}
\end{document}